# Unravelling the Antimicrobial Action Mechanism of Ribosomal Protein S30

J. Bhatt Mitra[1], V. K. Sharma[2,3*], M. Kumar[3,4], V. Garcia Sakai[5] A. Mukherjee[1,3*]

[1]Radiopharmaceuticals Division, Bhabha Atomic Research Centre, Mumbai 400085, India

[2]Solid state physics Division, Bhabha Atomic Research Centre, Mumbai 400085, India

[3]Homi Bhabha National Institute, Anushaktinagar, Mumbai-400094, India

[4]Radiation Biology & Health Sciences Division, Bhabha Atomic Research Centre, Mumbai 400085, India

[5]ISIS Neutron and Muon Source, Science and Technology Facilities Council, Rutherford Appleton Laboratory, Didcot, U.K

*Abstract*

Ribosomal protein S30 (RS30) exhibits potent antimicrobial activity against a broad spectrum of bacteria. Despite its efficacy, the underlying action mechanism remained elusive. In this study, we unravel the fundamental mechanism by which RS30 exerts its bactericidal effects, using a combination of microbiological assays and advanced biophysical techniques. Microbiological analyses reveal that RS30 kills bacteria primarily through membrane depolarization, despite limited membrane permeabilization, indicating an unconventional mode of action involving no or partial lysis of the membrane. Importantly, RS30 demonstrates time-dependent bactericidal activity with no detectable cytotoxicity toward mammalian cells, underscoring its high selectivity. This selective action was further confirmed using biophysical experiments on model membrane systems composed of anionic (bacterial mimic) and zwitterionic (mammalian mimic) phospholipids. Our measurements suggested that RS30 preferentially binds to anionic membranes via electrostatic interactions, undergoes a conformational transition from a random coil to an α-helix upon binding, and induces vesicle aggregation. Quasielastic neutron scattering (QENS) measurements provide microscopic insights, showing that RS30 significantly restricts the lateral diffusion of anionic lipids, thereby perturbing membrane dynamics and increasing susceptibility to external stress. Together, our findings uncover important insights into the antimicrobial action mechanism of RS30, characterized by selective membrane interaction, structural transformation, and dynamic modulation of lipid membranes.

**Corresponding authors:** sharmavk@barc.gov.in (V.K.S.); archanas@barc.gov.in (A.M.)

# 1. INTRODUCTION

Ribosomal protein S 30 (RS30), also known as ubiquicidin (UBI), is an essential component of the 40S subunit of the eukaryotic ribosome, playing a crucial role in translation. Unlike most ribosomal proteins, which are directly translated as unprocessed primary products of their mRNAs, RS30 is produced through the post-translational processing of a 133-residue Fau, protein. This processing generates two distinct products: a ubiquitin-like domain (FUb) from the N-terminus, and a 59-residue ribosomal protein, RS30, from the C-terminus [1, 2, 3,4]. RS30/UBI is conserved across eukaryotes, from yeast to humans; however, its production as part of a fusion protein is evolutionarily restricted to metazoans [1]. This restriction is largely attributed to the absence of the specific protease activity required to cleave Fau proteins in non-metazoan organisms [1]. RS30/UBI was first isolated from the cytosol of mouse macrophages (RAW264.7) by Hiemstra *et al.,* who named it ubiquicidin due to its sequence identity (38%) with full-length ubiquitin [5]. This peptide demonstrated potent antimicrobial activity, targeting both intracellular pathogens such as *L. monocytogenes* and *S. typhimurium* and extracellular pathogens including *S. aureus* and *E. coli* [5-7]. Additionally, RS30 has shown efficacy against methicillin-resistant *Staphylococcus aureus* (MRSA), supporting its role in combating drug resistant bacteria [8]. These findings underscore the dual functionality of RS30: as a ribosomal constituent essential for translation and as an antimicrobial peptide (AMP) in combating pathogens and mediating the host innate immune response. This versatility aligns with the broader observation, that many ribosomal proteins assume extra ribosomal roles [9].

The 2022 report from the Global Antimicrobial Resistance and Use Surveillance System (GLASS) highlights troubling trends in global antibiotic resistance, including a significant rise in resistance among common bacterial pathogens [11]. Of particular concern is resistance in *S. aureus*, which complicates treatment, leading to prolonged hospital stays and higher mortality rates [10]. The U.S. Centers for Disease Control and Prevention (CDC) has classified drug-resistant *S. aureus* infections among the top 18 antimicrobial drug-resistant threats [12]. While conventional antibiotics often lose effectiveness due to simple point mutations in bacterial targets, AMPs offer a promising alternative [13, 14]. These peptides primarily target the bacterial cell membrane through a complex mechanism, making resistance development less frequent [15, 16, 17].

Early diagnosis also plays a crucial role in combating antimicrobial resistance by reducing unwarranted antibiotic use. Single Photon Emission Computed Tomography (SPECT) and Positron Emission Tomography (PET) radiotracers targeting various molecular targets on bacteria have been successful in identifying deep seated focal infections [12, 18]. AMP-based agents have the potential to diagnose resistant bacteria and offer an advantage over molecules like antibiotics, which are limited by the emergence of resistance. Radiotracers labeled with $^{99m}$Tc and $^{68}$Ga, developed from RS30/UBI-derived fragments, have already shown promise as infection imaging probes for nuclear imaging of *S. aureus*-driven focal infections in both preclinical and clinical settings [8, 19-21]. Despite its promising potential in diagnostics and antibacterial therapy, the precise mechanism behind the interaction of RS30 with bacteria remains elusive; this cationic peptide can be envisioned to target the anionic bacterial cell membrane, which may serve as the primary site of its antimicrobial action.

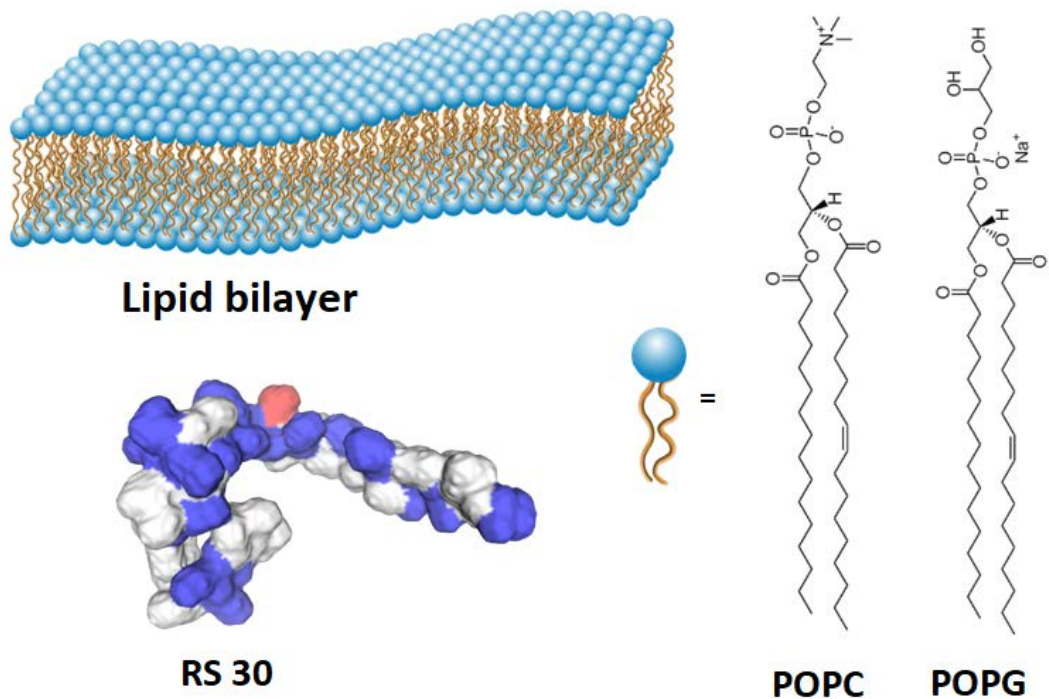

Figure 1: Schematic of a lipid bilayer and chemical structures of the lipids used in this study. Cartoon depicting the structure of RS30 modeled (SWISS-MODEL) from PDB ID: 6Y2L.72 (3 Å). Blue patches represent positively charged residues, while red patches denote anionic ones.

RS30 is rich in positively charged residues (13 lysines and 7 arginines; pI = 12.15), which likely facilitate its interaction with bacteria by engaging negatively charged membrane phospholipids. Bacterial membranes predominantly consist of anionic phospholipids, such as phosphatidylglycerol (PG), phosphatidylserine (PS), and cardiolipin (CL) [19]. For instance, PG and its lysylated derivative constitute ~ 95% of total phospholipids in the *S. aureus* membrane [22]. In contrast, the outer leaflet of mammalian membranes primarily comprises zwitterionic phospholipids, such as phosphatidylethanolamine (PE), phosphatidylcholine (PC), or sphingomyelin (SM), along with cholesterol, which mitigates the activity of AMPs [22, 23]. This variation in lipid composition between bacterial and mammalian membranes may play an important role in the selective interaction of AMPs with bacteria.

Here, we present the first report on the action mechanism of RS30 on bacterial membrane by, utilizing a combination of microbiological assays and biophysical techniques. RS30 shows both bacteriostatic and bactericidal impacts on *S. aureus*, consistent with the literature [5]. Additionally, RS30 is non-toxic towards mammalian cells, highlighting its selective antibacterial action. Moreover, flow cytometry measurements revealed that RS30 causes permeabilization and depolarization of the bacterial cell membrane to kill the bacteria, providing evidence that the membrane is a target for its action. AMPs are known to interact directly with phospholipid membranes without requiring protein receptors, allowing phospholipid membranes to serve as ideal model systems for understanding their action mechanism. We employed two model membrane systems composed from anionic and zwitterionic phospholipids to mimic bacterial and mammalian cell membranes, respectively. A series of biophysical experiments was conducted to uncover the physical basis of the selectivity of RS30 towards bacterial membranes. Dye leakage studies on model membranes, supported by bacterial assays, showed that direct interaction of RS30 with model bacterial membranes is sufficient to induce permeabilization and antimicrobial action. Isothermal Titration Calorimetry (ITC), Circular Dichroism (CD) and Dynamic Light Scattering (DLS) data confirmed that RS30 selectively interacts with the model bacterial membrane. These findings highlight the crucial role of electrostatic interactions in RS30's selective binding to bacterial membranes. The peptide's highly cationic nature enables strong interactions with the anionic phospholipids of bacterial membranes, resulting in membrane rupture, depolarization, and ultimately bacterial death. Furthermore, Quasielastic Neutron Scattering (QENS) data revealed that RS30 restricts lipid lateral diffusion, causing membrane strain. This strain makes bacterial membranes more vulnerable to local perturbations and stress, ultimately leading to membrane rupture, as observed in dye leakage assays. This study reveals that the bacterial membrane serves as a critical target for the antimicrobial action of RS30. The findings of this study provide the first insights into the RS30-membrane interaction, potentially advancing and enhancing the theranostic application of this AMP. Additionally, it sheds light on how positively charged proteins, such as histones and other ribosomal proteins, may assume an additional antimicrobial role.

## 2. MATERIALS AND METHODS

**Materials**

*S aureus* ATCC 25923 was procured from Microbiologics (MN, USA). Mueller Hinton broth (MHB), Resazurin, Agar, Dulbecco's Modified Eagle Medium (DMEM), Fetal Bovine Serum (FBS) and 3-(4,5-dimethylthiazol-2-yl)-2,5-diphenyltetrazolium bromide (MTT) were obtained from HiMedia Labs (Mumbai, India). RS30 was purchased from ABI scientific (VA, USA). Propidium iodide (PI), Melittin, Dimethyl Sulfoxide (DMSO), Calcein and Triton X-100 were obtained from Sigma Aldrich (MO, USA). Carbonyl cyanide m-chlorophenyl hydrazine (CCCP) and 3,3-Diethyloxacarbocyanine iodide or DiOC2(3) were supplied as parts of bacterial membrane potential kit procured from Thermo Fisher Scientific (MA, USA). Human embryonic kidney (HEK 293) cells were procured from the National Centre for Cell Sciences (NCCS) located in Pune, India. All phospholipids used in this study were obtained from Avanti Polar Lipids (AL, USA).

**Determination of minimum inhibitory concentration (MIC) and bactericidal concentration (MBC)**

A single colony of *S. aureus* ATCC 25923 , was inoculated in MHB and incubated overnight at 120 rpm and 37 °C. The experiment was conducted in a 96 microtiter plate using a broth dilution assay. RS30 was serially diluted (twofold per step) in M9 media supplemented with 0.01% Tryptone. The cells were washed twice with the media and diluted to a final concentration of ~$10^5$ colony forming units (CFU) per well. The final volume in each well was 200 µL. To determine the minimum bactericidal concentration (MBC), 5 µL aliquots were taken from the wells and spotted onto MHB-agar plates. The plates were incubated overnight at 37 °C. After 24 hours, 30 µL of 0.75 mg/mL resazurin dye was added to each well and incubated overnight at 37

°C and 150 rpm. For the time kill assay, *S. aureus* was cultured and washed with media as described before. The experiment was carried out in microfuge tubes. Approximately, $10^6$ cells were incubated with the minimum inhibitory concentration (MIC) of RS30 (2 µM) in a final volume of 500 µL adjusted with M9 media. The cells and RS30 were incubated in media at 37 °C and 120 rpm. At the required time intervals, 50 µL aliquots were taken from each reaction mixture and appropriate dilutions were plated onto MHB agar plates, followed by incubation at 37 °C for 16 h. The experiment was performed in duplicates and repeated three times. The surviving fractions were plotted against time (min). The data was analyzed using Origin Graphing and Analysis 6.0 Pro software.

**Flow cytometry**

Approximately, $10^7$ CFU/mL, log phase *S.aureus* were washed with media and incubated with RS30 at four times MIC for 1 h at 37 °C and 120 rpm. To study permeabilization, RS30 treated cells were washed with saline followed by staining with DNA stain, PI at final concentration of 45 µM. Incubation with PI was carried out for 30 min at 37 °C and 120 rpm. Isopropanol served as a positive control for membrane permeabilization. To assess depolarization of *S. aureus* membranes, RS30 treated bacteria were stained with membrane potential sensing dye DiOC2(3) at a final concentration of 30 µM for 30 min at 37 °C and 120 rpm. A positive control was prepared by incubating bacteria under similar conditions with 5 µM of CCCP. All samples were analyzed by Guava EasyCyte flow cytometer (Darmstadt, Germany).

**Haemolysis assay**

Fresh human blood samples were obtained from healthy donors with the approval of the Institutional Medical Ethics Committee. Erythrocytes were isolated from heparinised human blood by centrifugation at 2000 g for 5 min after washing three times with PBS at 4°C. After removing the supernatant, the cell pellet was resuspended in 2 mL of Phosphate Buffered Saline (PBS). The erythrocyte suspension was incubated with various concentrations of peptides for 30 min at 37°C at a final volume of 200 µL in microfuge tubes in duplicates at a dilution of 4 % (v/v). Samples were centrifuged at 2000 g for 5 min, and the absorbance of the supernatants was measured at 540 nm using a polarstar omega plate reader from BMG Labtech (Ortenberg, Germany). Erythrocytes lysed with 0.1 % triton X-100 and melittin (17 µM) were used as positive controls. Percentage haemolysis was calculated using the formula:

$$\text{Haemolysis (\%)} = \frac{(A_{SAMPLE} - A_{NC})}{(A_C - A_{NC})} \times 100 \tag{1}$$

Where $A_{SAMPLE}$=Absorbance recorded for samples at 540 nm, $A_{NC}$ = Absorbance recorded for untreated (negative control) erythrocytes at 540 nm, $A_C$ = Absorbance recorded for Triton X-100 treated (positive control) erythrocytes at 540 nm.

**Cytotoxicity assay**

Human embryonic kidney (HEK 293) cells were maintained in DMEM supplemented with 10% FBS and 1% antibiotic-antimycotic solution at 37 °C, 5% $CO_2$ in humidified atmosphere. Cell cytotoxicity was measured by performing MTT assay. Briefly, 2000 cells/100 µL were seeded in a 96-well plate and kept overnight in incubator at 37 °C, 5% $CO_2$. RS30 was added to the wells and incubated for 24 h. Melittin (17 µM) was used as a positive control for cytotoxicity assay. Following incubation, 10 µL of MTT (5 mg/mL) was added to each well and after 3 h the supernatant was removed and 100 µL of dimethyl sulfoxide (DMSO) was added. After 1 h incubation, the absorbance was recorded at 570 nm. One-way ANOVA was used to establish the presence of significant groups, and Tukey's test was subsequently utilized to conduct the comparison of means.

**Preparation of Unilamellar Vesicles (ULVs)**

Two types of ULVs, namely large unilameller vesicles (LUVs) and small unilamellar vesicles (SUVs), were prepared based on the experimental requirements. The negatively charged phospholipid, i.e. 1-palmitoyl-2-oleoyl-sn-glycero-3-phosphatidylglycerol (POPG), was used to prepare ULVs representing bacterial model membranes, whereas zwitterionic phospholipid, 1-palmitoyl-2-oleoyl-sn-glycero-3-phosphocholine (POPC) served as mammalian membrane model. LUVs were formulated using the extrusion method, as previously described by us [24]. Briefly, POPC and POPG were dissolved in chloroform or a mixture of chloroform and methanol (7:3, v/v), respectively. Thin films of POPG and POPC were obtained by gently evaporating the respective organic solvents first under $N_2$ stream followed by complete evaporation under vacuum. The films were then hydrated in PBS and subjected to freeze-thaw cycles (10 times) to obtain multilamellar vesicles (MLVs). To prepare LUVs, the MLVs were extruded using a mini extruder from Avanti Polar Lipids Inc., passing them through polycarbonate filters with a pore size of 0.1 µm. For SUVs, MLVs underwent sonication using the, Q700CA sonicator from Qsonica Sonicaters (Illinois,USA) . The sonication process involved cycles of 2.5 minutes, with 30 seconds of sonication followed by 30 seconds of rest, at amplitude of 50%. As will be demonstrated later, the incorporation of RS30 induces aggregation of anionic POPG vesicles. However, this aggregation can be effectively prevented by incorporating 3 mol% 1,2-dipalmitoyl-sn-glycero-3-phosphoethanolamine-N-(methoxy(polyethylene glycol)-2000 (PEG-PE) during vesicle preparation. ITC further confirmed that the presence of 3 mol% PEG-PE does not significantly alter the interaction between RS30 and the POPG membranes (Fig. S1). Therefore, to ensure vesicle stability and avoid aggregation during CD and QENS experiments, POPG vesicles were prepared with 3 mol% PEG-PE. Phospholipid concentrations were determined based on dry weight, while peptide concentrations were quantified by measuring absorbance at 280 nm.

**Dynamic light scattering (DLS)**

Dynamic light scattering (DLS) experiments were conducted to examine the distribution of hydrodynamic size of LUVs at varying peptide to lipid molar ratios (% P/L). Large micron size impurities were removed from the buffer by filtration through 0.22 µ filter. The initial concentration of LUVs used was 650 µM. The experiments were performed in PBS at pH 7.4, utilizing a Zetasizer nano zs system from Malvern instruments. A 633 nm He-Ne laser served as the light source, and the scattered light was measured at an angle of 173 $^\circ$. The temperature was consistently maintained at 25 °C throughout the duration of the experiment.

**Isothermal Titration Calorimetry (ITC)**

ITC experiments were conducted using the MicroCal iTC200 system from Malvern instruments. The calorimeter cell, with a volume of 200 µL, was loaded with either anionic (POPG) or zwitterionic (POPC) LUVs at a concentration of 650 µM. The syringe was filled with RS30 (250 µM) for the titrations. All titrations were performed until the heat signal reached saturation. To minimize the impact of equilibration artefacts, the first injection of 0.4 µL volume was discarded from the analysis, as these injections are often associated with equilibration issues [25]. The titrations were carried out at a temperature of 25 °C. The obtained isotherms from the interaction between the peptide and POPG LUVs were fitted with a single-site binding model using the ITC 200 software. This model assumes the presence of independent binding sites for the ligand, providing the stoichiometry ($n$), enthalpy change ($\Delta H$), and association constant ($K$) for their respective interactions [25,26].

**Dye Release Assay**

To study membrane disruption, a dye release assay was performed. Initially, phospholipid films (POPC and POPG) were hydrated with a green fluorescent dye calcein at a concentration of 70 mM. The hydrated films were then extruded through a 0.1 µm polycarbonate filter to prepare LUVs containing the dye. These dye-loaded LUVs were subsequently exposed to peptides at varying concentrations in a 96-well plate format. The release of the dye over time was monitored by measuring the increase in fluorescence resulting from the interaction between the peptide and the LUVs. This fluorescence measurement was carried out at room temperature using a plate reader (Polarstar Omega, BMG Labtech, Offenburg, Germany). As positive controls for complete dye release ($F_{100}$), Triton X-100 treated cells were used, while untreated cells were considered to have 0% dye release ($F_0$). The percentage of normalized dye release was calculated using the following formula:

$$\text{Dye release (\%)} = \frac{(F-F_0)}{(F_{100}-F_0)} \times 100 \qquad (2)$$

**Circular Dichroism (CD)**

The conformation of RS30 in aqueous as well as hydrophobic environment was studied by CD spectroscopy in the far ultraviolet region (Far-UV). The concentration of samples was standardized based on signal intensity and high voltage. Experiments were performed on a J-815 CD spectrometer from Jasco (Tokyo, Japan). Samples were either prepared in deionized water or with increasing percentage (*v/v*) of trifluoroethanol (TFE) which was used as a solvent to observe changes in conformation of the peptide in the membrane like environment. Peptide conformation was also studied with SUVs at % P/L= 4 %. Background corrected CD signals were recorded for samples at wavelengths 260-190 nm (Far-UV). Spectra were collected at 25 °C using a path-length of 1 mm. CD signal (or ellipticity) in milldegrees was converted to mean residual ellipticity (MRE) represented as $\theta_m$ based on Eq.S1 in the supporting information.

**Quasielastic Neutron Scattering (QENS)**

To investigate the effects of RS30 on membrane dynamics, QENS experiments were carried out on anionic 50 mM POPG +3 mol % PEG-PE LUVs (based on $D_2O$) in the absence and presence of 2 mol % RS30 at 27° C. The experiments were performed using the near-bakcscattering neutron IRIS spectrometer at the ISIS facility in the UK. The spectrometer utilized the pyrolytic graphite (002) analyzer in the offset mode, providing an energy resolution of 17 µeV (full width at half-maximum). The energy transfer range was set between 0.3 and 1.0 meV, while the wave-vector transfer ($Q$) range covered 0.5 to 1.8 Å$^{-1}$. For the QENS measurements, the samples were placed in annular aluminium sample holders with 0.5 mm internal spacing to minimize multiple scattering effects. QENS spectra were also recorded for the solvent ($D_2O$) for reference. For instrument resolution, QENS measurements were performed on standard vanadium. The data reduction, including detector efficiency correction and background subtraction, was carried out using MANTID software [27].

**RESULTS AND DISCUSSION**

*RS30 shows time dependent bactericidal activity without exerting cytotoxic effects on mammalian cells*

Effect of RS30 on bacterial growth was evaluated by a resazurin (alamar blue) reduction assay in a microtiter plate format (Fig. 2). Metabolically active cells reduce resazurin dye (blue colour) to pink resafurin. Therefore, MIC was measured by visually observing the disappearance of pink color in the wells. The concentration of peptide which caused the disappearance of bacterial colonies in the spot assay was determined as MBC. Both MIC and MBC were found to be 2 µM indicating that RS30 had bacteriostatic as well as potent bactericidal impact on *S. aureus*. Furthermore, RS30 killed *S. aureus* in a time dependent manner with complete elimination of *S. aureus* colonies within 30 minutes of incubation. Furthermore, the therapeutic potential of the peptide was assessed by determining its toxicity on RBCs and Human Embryonic kidney cells (HEK 293) by haemolysis and cytotoxicity assays respectively. It was observed that at tested concentrations, RS30 induced no haemolysis while Melittin and Triton X-100 (positive controls) were haemolytic to erythrocytes (Fig. 2c). The cell cytotoxicity data shows that the treatment of HEK 293 (immortalized non-cancer cells) with RS30 induced minimal cytotoxicity (Fig. 2d) whereas, Melittin (positive control) was found to be cytotoxic to HEK 293 cells. In summary, RS30 was non-haemolytic and exhibited minimal cytotoxicity to mammalian cells at the tested concentrations, demonstrating its potential for theranostic applications.

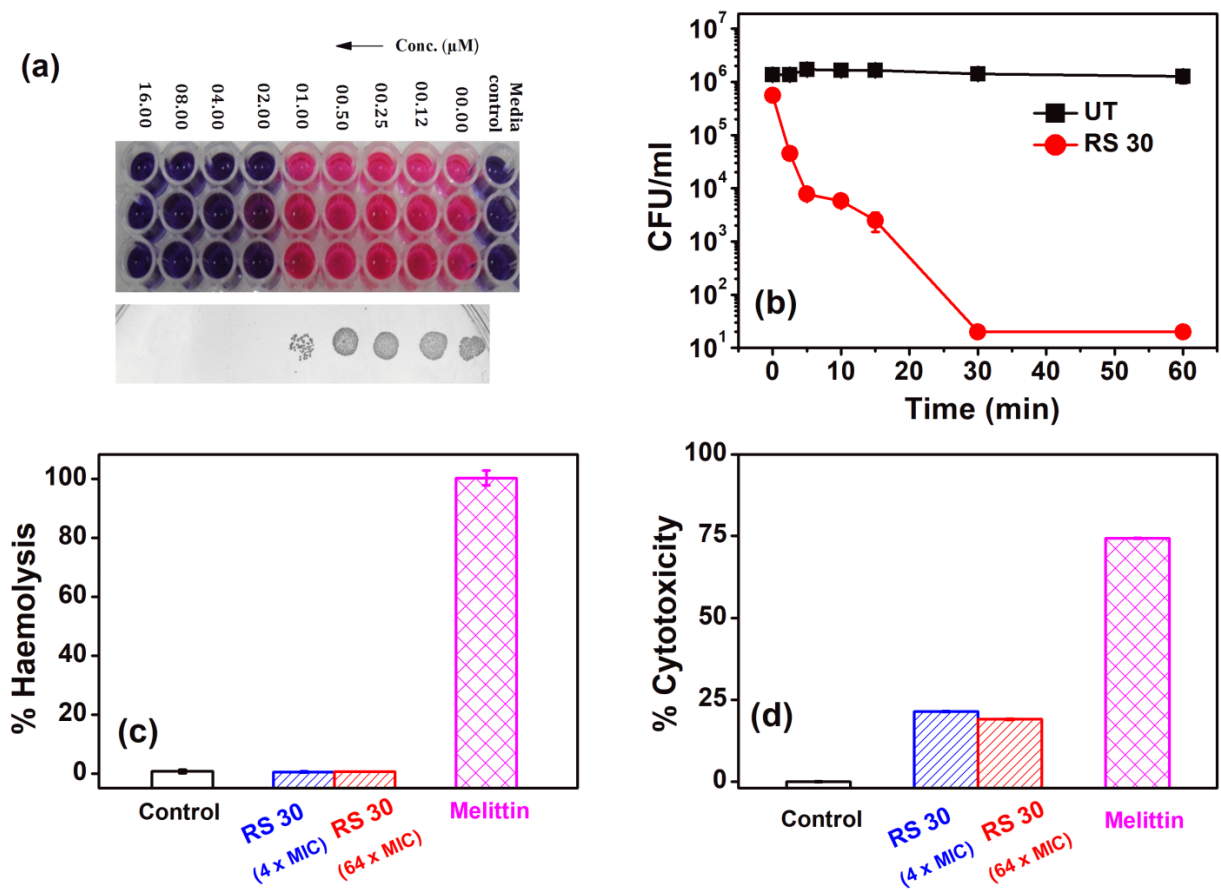

Figure 2: (a) MIC and MBC of RS30 against *S. Aureus*. (b) Viability of *S. aureus* in a time dependent manner at MIC. Data for untreated (UT) bacteria is shown for reference. (c) Haemolysis and (d) cytotoxicity of RS30 against mammalian cells (HEK 293). Untreated RBCs and HEK 293 cells are used as controls for haemolysis and cytotoxicity assays, respectively. Melittin was used at a concentration of 17 µM as positive control.

### RS30 kills S. aureus by depolarization and permeabilization of the membrane

A well-known mechanism employed by AMPs to kill bacterial cells is permeabilization of membrane followed by its depolarization [28-32]. To explore if RS30 utilized this strategy to adversely affect the survival of *S. aureus*, permeabilization and depolarization of *S. aureus* membrane was studied by flow cytometry. DiOC2(3) is a green carbocyanine dye that senses membrane potential by aggregating in cells with higher membrane potential (live cells) and emitting red light. A high red to green ratio indicates a higher relative membrane potential. CCCP was used as a positive control while measuring relative membrane potential of cells. It is a proton-ionophore which increases the proton permeability of the membrane leading to its depolarization. It was found that full length RS30 severely depolarized the *S. aureus* (Fig.3) membrane when compared to the CCCP control.

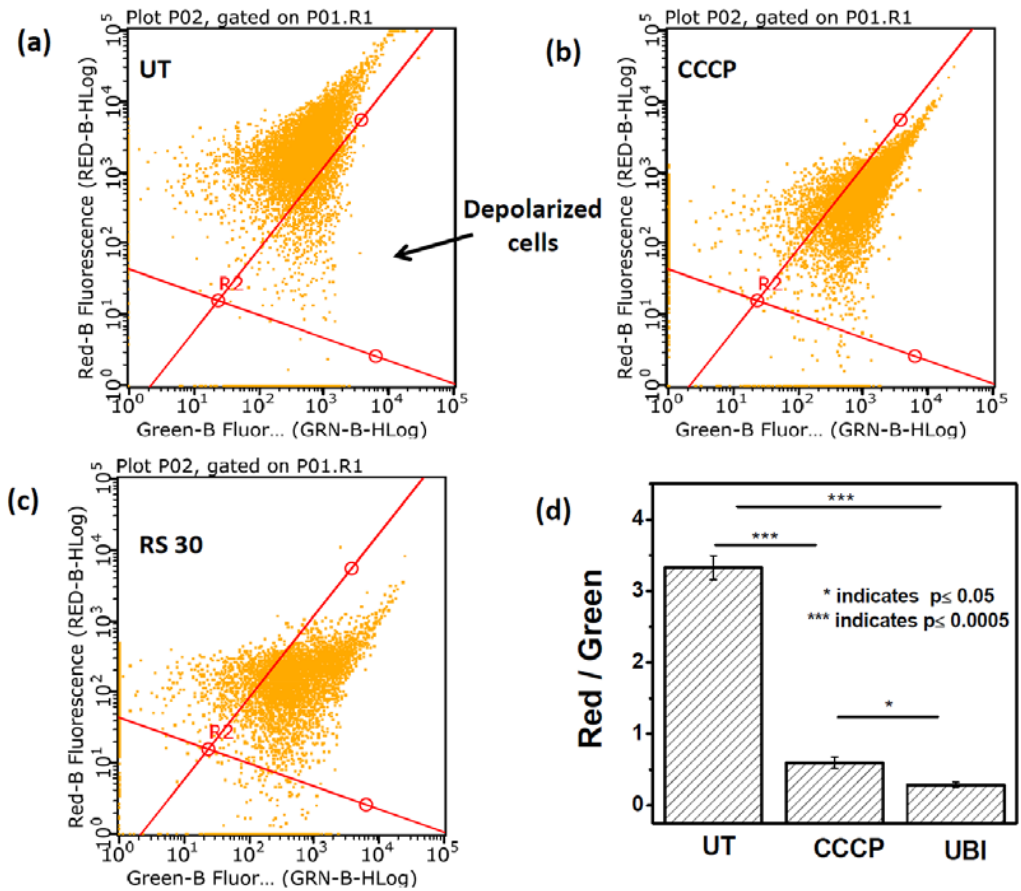

Figure 3: Dot plots showing live (high red) and depolarized *S. aureus* (high green) under different treatments: (a) untreated (UT), (b) CCCP (positive control) and (c) RS30 (8 µM). (d) Bar graph depicts ratio of red and green fluorescence intensities of bacterial cells with designated treatments. Statistically significant groups are also marked.

Further, permeabilization of *S. aureus* membrane was studied by following the uptake of a membrane impermeant, red fluorescent dye known as PI. The uptake of PI by bacterial cells denotes a compromised membrane barrier. Our results indicate that RS30 causes minimal permeabilization of the *S. aureus* membrane (~14% normalized permeabilization) compared to the isopropanol control, even at four times the MIC concentration (Fig. 4).Interestingly, despite this limited permeabilization, RS30 induces severe membrane depolarization (Fig. 3), surpassing that caused by the CCCP control. This is counterintuitive, as membrane depolarization in most AMPs typically follows significant membrane permeabilization. These findings suggest that the mechanism of action of RS30 is fundamentally different from that of conventional AMPs. The observed results can be explained by mechanisms such as transient pore formation and/or membrane thinning [33-36]. Transient water pores may form in response to ionic charge imbalances across the membrane. These metastable pores can briefly allow ion passage, leading to a rapid reduction in transmembrane potential. As the potential diminishes, the driving force for pore formation weakens, resulting in pore closure, thus enabling membrane depolarization without sustained permeabilization. RS30 may also induce membrane thinning, which reduces the packing density of lipids and generates nanoscale cavities. These structural perturbations lower the translocation free energy for water molecules across the hydrophobic core of the membrane, further contributing to depolarization without causing extensive permeabilization.

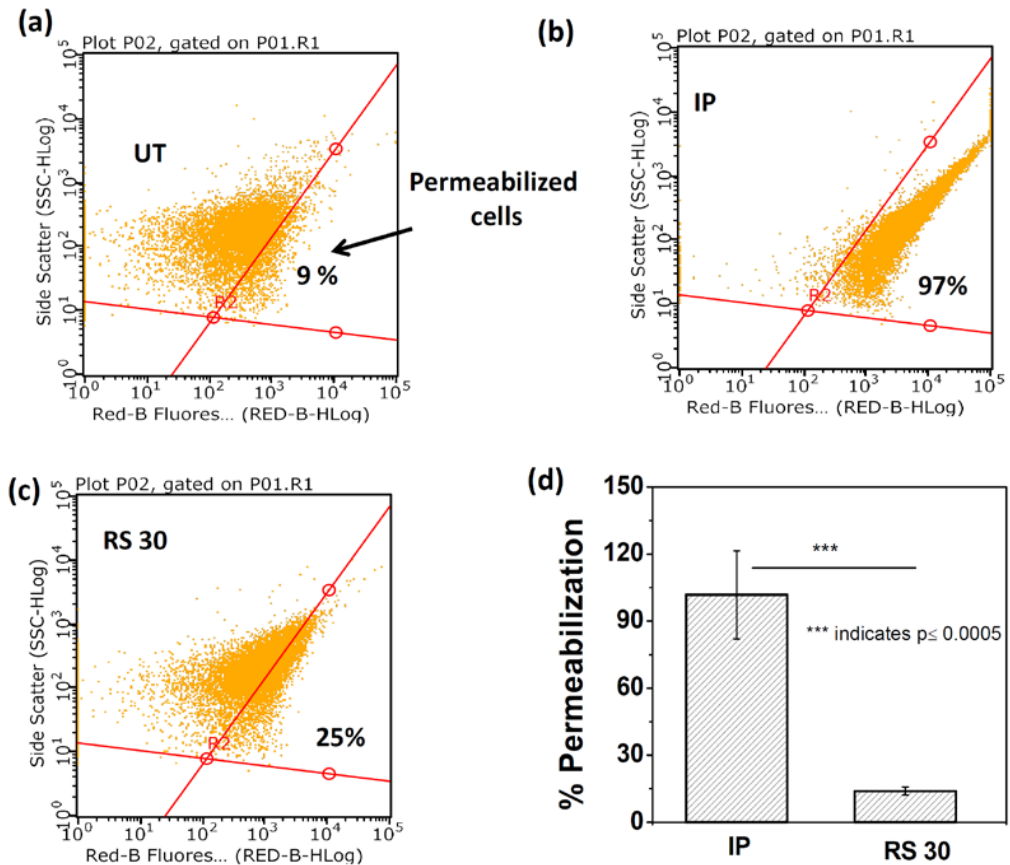

Figure 4: Dot plots showing % permeabilized *S. aureus* under different treatments: (a) untreated (UT), (b) isopropanol (IP) (positive control), (c) RS30 (8 µM). (d) Bar graph shows normalized permeabilization of samples.

*RS30 induces selective aggregation of anionic vesicles mimicking bacterial membranes*

Surface binding of cationic AMPs with anionic LUVs can potentially facilitate aggregation or fusion of anionic vesicles due to the screening of Coulomb repulsion. If the size of LUV after aggregation increases to the order of a few micrometers, the vesicle solution should transform from clear to turbid. To observe this transformation, transmission measurements of light (500 nm) through anionic POPG LUVs as a function of concentrations of RS30 were performed. These measurements show that this peptide induced aggregation in anionic LUVs (Fig. 5). The reduction in the transmission of light occurred with increasing % P/L (peptide to lipid) values. As RS30 possesses a high positive charge (+19), it screens repulsive interactions between the anionic vesicles effectively. It is worth noting that the aggregation of anionic vesicles has been reported for several antimicrobial peptides and is attributed to charge neutralization of vesicle surface [24, 37-39]. The energy released upon peptide-membrane interaction and conformational change in peptide drives the aggregation of vesicles in these cases [39]. Our DLS data supports these findings, confirming the increase in the average diameter of anionic LUVs with rising peptide concentrations. In contrast, the size of zwitterionic LUVs (POPC) remains unchanged. Incorporation of PE-PEG, which is known to prevent vesicle aggregation through steric stabilization, effectively abrogated the peptide-induced changes in the size of anionic LUVs when added at 3 mol %, as shown in Fig. S1 of the supporting information [38, 40]. This further corroborates the aggregation of anionic vesicles by RS30 and the critical role of electrostatic interactions in driving this process.

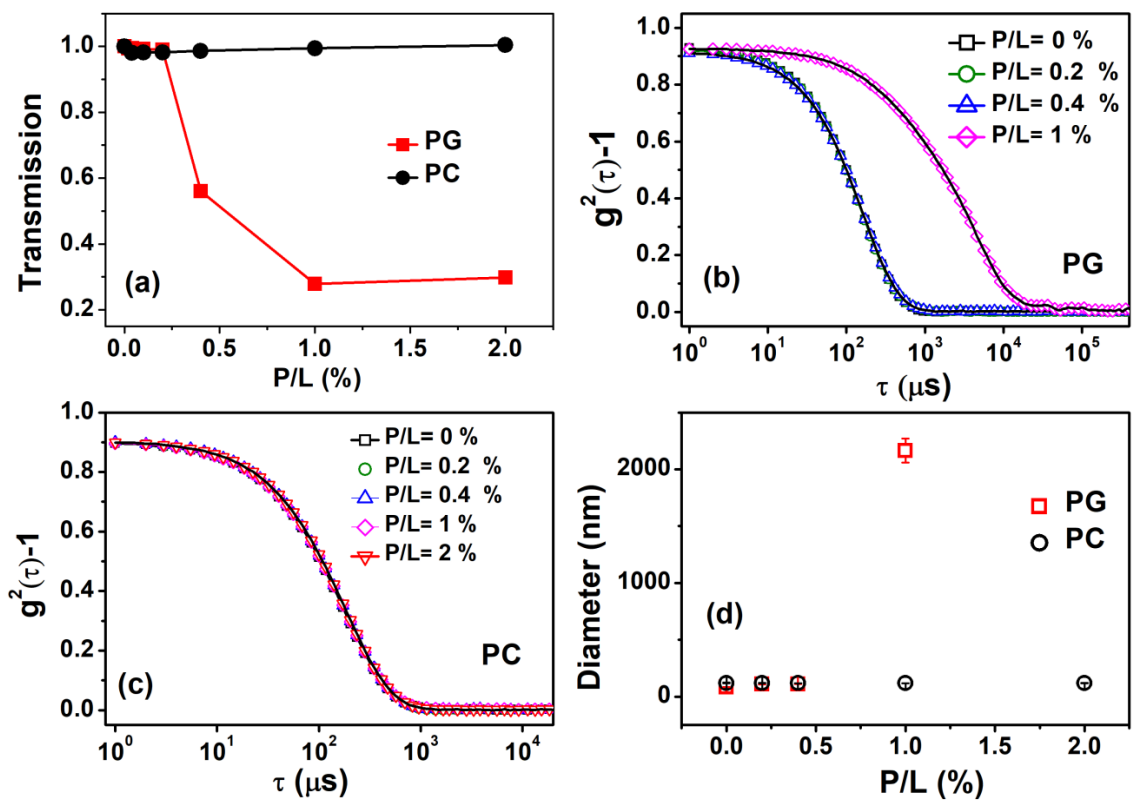

Figure 5: (a) Transmission of light (500 nm) in 650 µM POPG and POPC LUVs with varying concentrations of RS30. Intensity auto-correlation functions for (b) POPG and (c) POPC LUVs at different P/L ratios. (d) Variations in *z*-average diameters of POPC and POPG LUVs with P/L.

*Thermodynamic aspect of membrane-RS30 interaction*

Isothermal Titration Calorimetry (ITC) experiments were conducted to investigate the thermodynamics of membrane-RS30 interactions. These experiments were performed until the binding sites on the anionic (POPG) LUVs were saturated as observed from ITC traces and isotherms (Fig 6). The results produced complex isotherms for peptide-lipid interactions (Fig. 6). For RS30 -POPG binding, the enthalpy ($\Delta H$) and association constant ($K$) were determined using single-site binding model (Eq. S5) and are found to be 5.5 ± 0.2 kcal/mol and (4.1±1.8)×10$^6$ M$^{-1}$, respectively. This association constant enables the calculation of the Gibbs free energy associated with the binding of RS30 to the lipid membrane, utilizing the relation $\Delta G = -RT \ln 55.5 K$, where 55.5 M denotes the concentration of water. The resulting $\Delta G$ for RS30 is found to be -11.4 kcal/mol. The endothermic nature of the interaction suggests a dominant contribution of entropy (T$\Delta S$=16.9 Kcal/mol), likely due to the aggregation of anionic vesicles by RS30, even at low peptide-to-lipid (P/L) ratios [41]. The binding stoichiometry (*n*) revealed that 42 POPG molecules interact with each peptide molecule. This high stoichiometry supports surface binding/adsorption of RS30 to phospholipid head groups on anionic LUVs rather than insertion into the membrane [37, 42].These findings align with the DLS data and are consistent with reports for other AMPs and membrane-binding peptides, such as melittin [37,42]. Such AMPs cause lethal membrane perturbations by interacting with phospholipid head groups, with the preference for head groups versus tails dictated by the ratio of positively charged to hydrophobic residues in the peptide [37,42].Titrations of RS30 into zwitterionic (POPC) LUVs (Fig. 6) yielded uniform molar heat values across various P/L ratios, indicating no significant membrane-peptide interactions. Thus, both DLS and ITC data confirm the selective interaction of RS30 with anionic membranes which is in agreement with our *in vitro* data. Preferential binding to anionic membranes underscores the importance of membrane charge in peptide-phospholipid interactions. Selectivity for bacterial cells over host cells is critical for reducing *in vivo* toxicity and achieving a high therapeutic index, which remains a major challenge for translating AMPs into clinical use. Moreover, selective binding is expected to enhance signal-to-noise ratios in SPECT or PET-based infection diagnostics using RS30/UBI-based tracers.

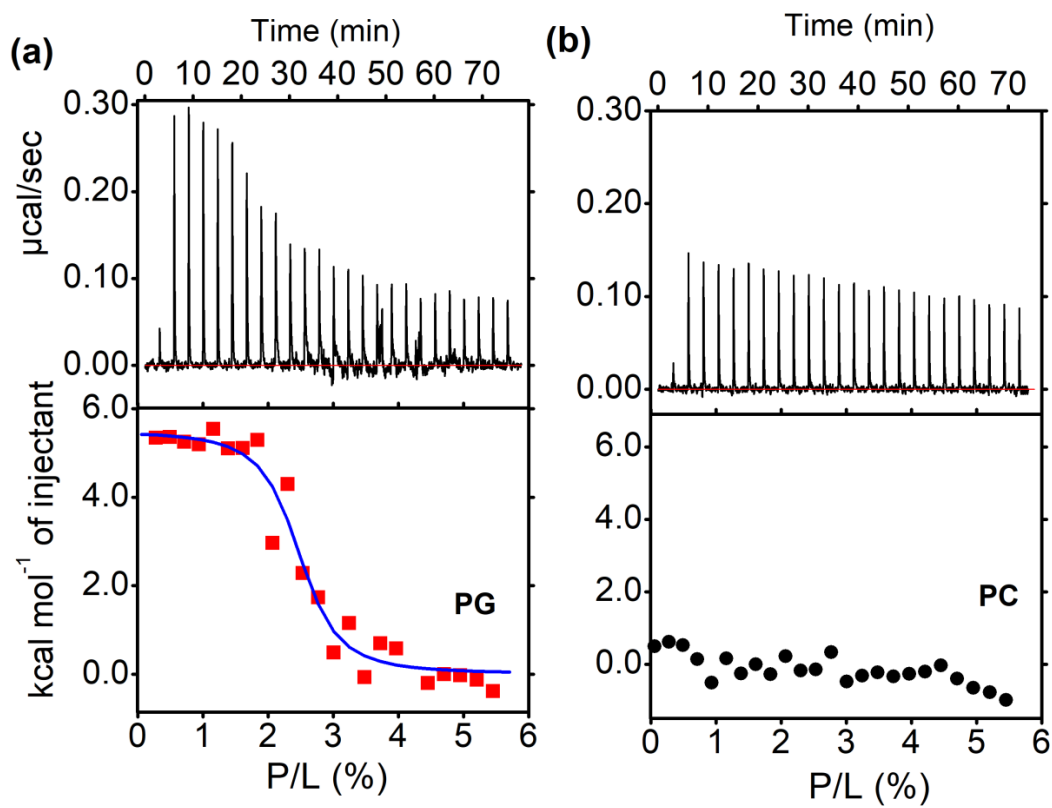

Figure 6: ITC traces and isotherms of RS30 with (a) POPG (anionic) (b) and POPC (zwitterionic) LUVs.

*RS30 selectively disrupts anionic phospholipid membranes*

To investigate whether the association of RS30 with anionic LUVs leads to membrane disruption, dye release assays were performed using calcein-loaded vesicles. LUVs encapsulating calcein, a green fluorescent dye, were prepared, and the percentage of dye release was monitored over time at various peptide concentrations. RS30 selectively induced dye leakage from anionic POPG LUVs in a concentration-dependent manner; however, even at concentrations tenfold higher than the MIC, the maximum calcein release remained below 50% (Fig. 7). In contrast, no dye leakage was observed from zwitterionic POPC LUVs under identical conditions. These observations highlight two key points: first, RS30 exhibits selective interaction and disruption of anionic membranes-representative of bacterial membranes-which aligns with findings from DLS and ITC studies. Second, the limited permeabilization of POPG LUVs even at concentration of ten times of MIC supports our proposed mechanism of action of RS 30, involving transient pore formation and/or membrane thinning rather than complete membrane lysis. This mechanism was also suggested by flow cytometry results on *S. aureus*, which revealed only partial membrane permeabilization, even at four times the MIC. We propose that electrostatic interactions drive the initial binding of RS30 to bacterial membranes, followed by the involvement of hydrophobic residues in mediating downstream disruptive effects.

*RS30-anionic membrane interaction promotes coil-helix transition in peptide*

As shown by transmission and DLS measurements, RS30 induced aggregation in anionic vesicles diminishes the CD signal due to scattering. Therefore, to study conformation of this peptide in the bacterial membrane model, POPG LUVs (anionic) were prepared with 3 mol % PE-PEG to inhibit peptide-induced vesicle aggregation [38]. The addition of PE-PEG to POPG retained interaction with the peptide while inhibiting the aggregation (Fig. S1, S.I.). Conformational changes, especially, coil-to-helix transition upon membrane binding, is a well-known phenomenon for a number of AMPs [29, 43]. TFE has often been used to provide a "membrane like environment" to probe these conformational changes [44-46]. Although trifluoroethanol (TFE) has an additional tendency to induce/stabilize the helical secondary structure in peptides, results from experiments performed in TFE tend to be in agreement with outcomes of studies done with lipid micelles or vesicles, which are physiologically closer to membranes [44, 46]. It is also useful due to the fact that LUVs scatter light, leading to loss of signal. Mean residue ellipticity or MRE ($\theta_m$) was calculated for RS30 at varying % of TFE (Eq.S1) and in POPG LUVs consisting of 3 mol % of PE-PEG, as shown in Fig. 8. It was observed that the peptide underwent conformational changes with the increasing amount of TFE (v/v). At 0 % (v/v) TFE, the peptide showed random coil conformation, as evident from minima around 190 nm. RS30/ clearly showed coil to-helix transition (minima at 222 and 208 nm) with an increase in the amount of TFE (v/v). It is found that helicity increases with an increasing amount of TFE (v/v) and it acquires a maximum value at 80% (v/v) TFE. In the membrane environment as well, it also underwent a conformational change as depicted in Fig.8 (lower panel).

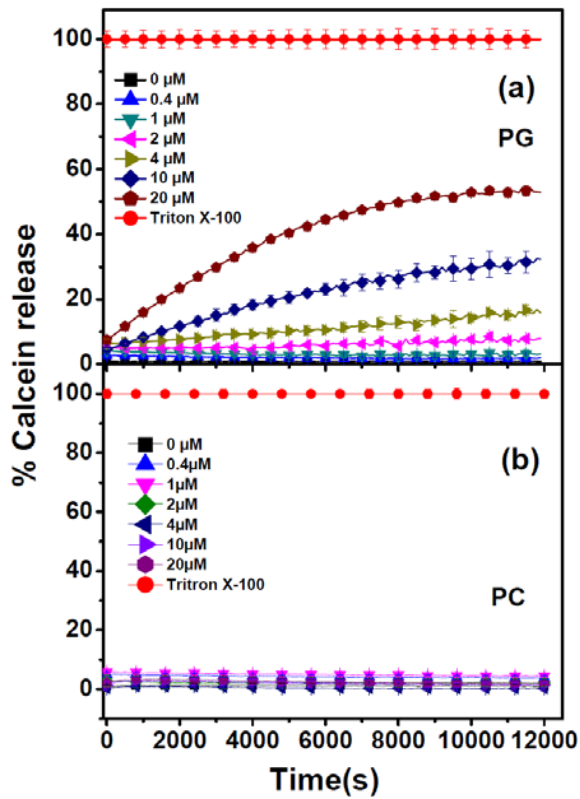

Figure 7: % Calcein release in (a) POPG and (b) POPC LUVs with time at varying concentrations of RS 30. Tritron X-100 was used as a positive control.

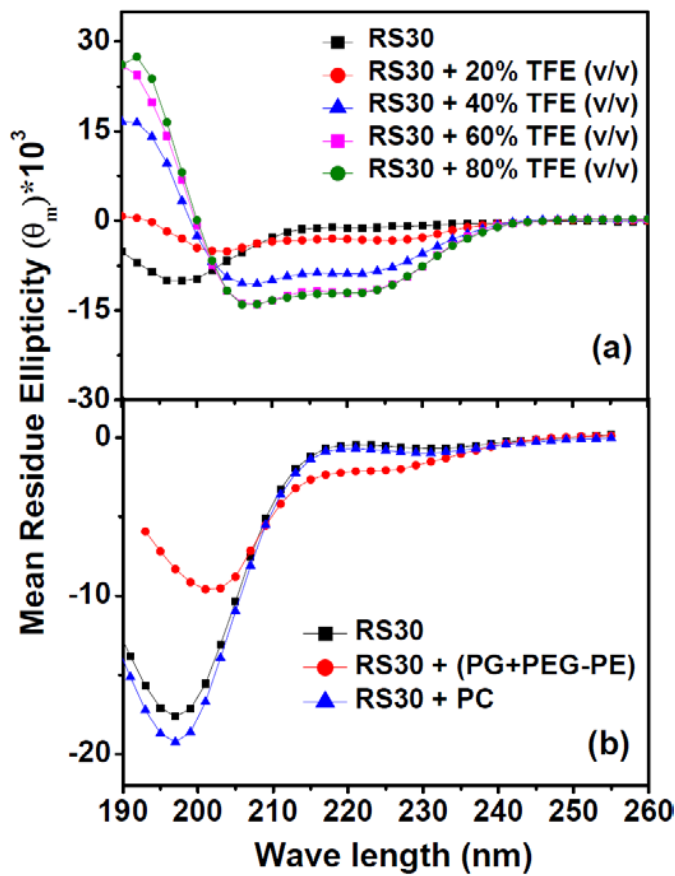

Figure 8: CD spectra of RS30 with (a) varying % of TFE (v/v) and (b) SUVs composed of POPG + PEG-PE (3 mol %) and POPC.

*RS30 restricts the lateral motion of phospholipid in anionic membranes*

So far, it has been shown that RS30 selectively interacts with the anionic membrane suggesting that electrostatic interactions play an important role in the peptide-membrane interactions. It is of great interest to understand the action mechanism, at the molecular level, by which RS30 interact with the anionic membrane. QENS is an excellent tool to investigate the membrane dynamics at the molecular scale [47-53]. QENS experiments were performed on POPG vesicles in the absence and presence of 2 mol% RS30. The scattered intensity from the POPG vesicles was obtained by subtracting the $D_2O$ contribution, as described in Eq. S(8). Significant quasielastic (QE) broadening is observed for the lipid membranes, indicative of the stochastic motions of lipid molecules (Fig. 9). Notably, the incorporation of RS30 results in a clear reduction in QE broadening, suggesting that RS30 modulates the membrane dynamics. Within the spatial and temporal resolution accessible by QENS, two distinct types of lipid motion are typically observed: (i) lateral diffusion within the membrane leaflet on the nanosecond timescale, and (ii) faster internal motions of the lipid chains on the picosecond timescale [47-49]. Assuming these motions are independent, the total scattering law for the lipid membrane can be modeled accordingly [see supporting information].

$$S_{mem}(Q,\omega) = [A(Q)L_{lat}(\Gamma_{lat},\omega) + (1-A(Q))L_{lat}(\Gamma_{lat}+\Gamma_{int},\omega)] \quad (3)$$

where $S_{mem}(Q,\omega)$ is the scattering law for membrane, $A(Q)$ is the Elastic Incoherent Structure Factor (EISF), $\Gamma_{lat}$ and $\Gamma_{int}$ are the HWHM's of the Lorentzian corresponding to lateral and internal motions of lipid respectively.

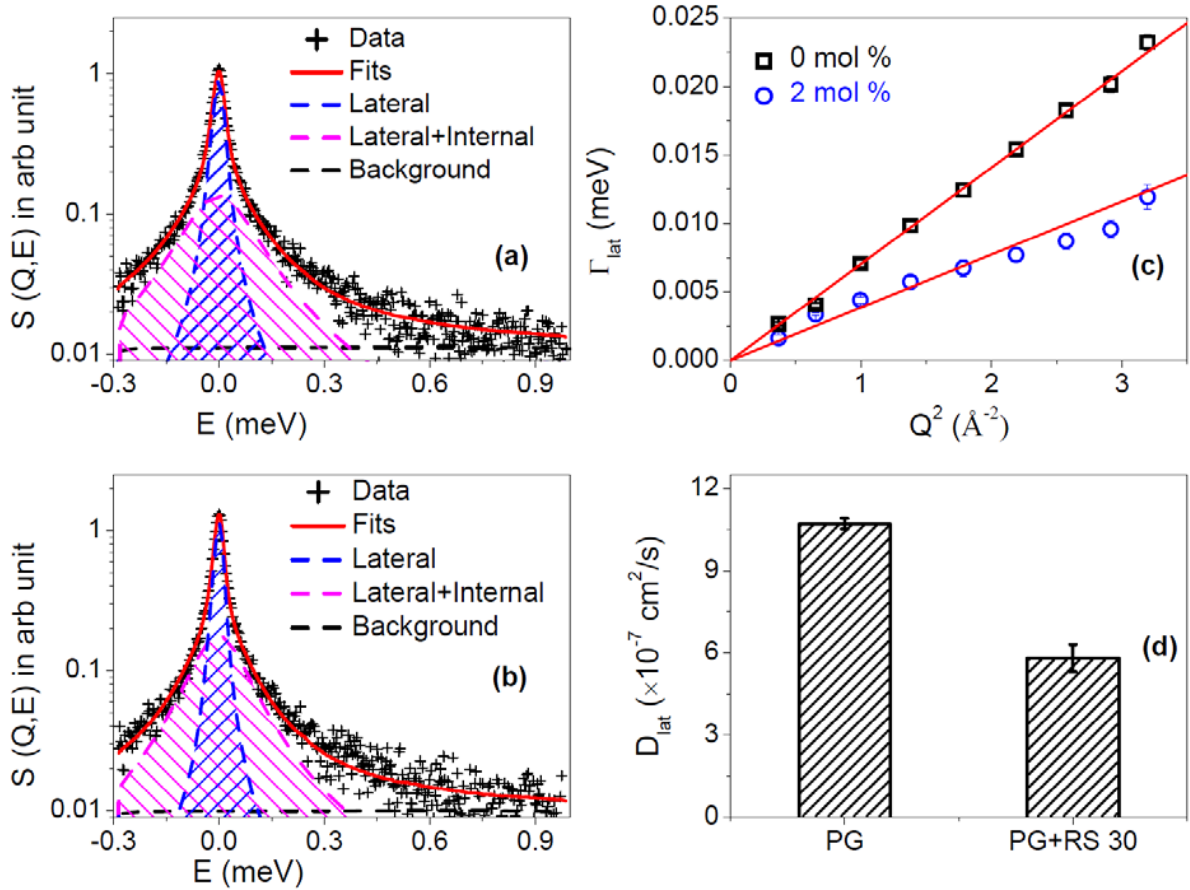

Figure 9: Typical fitted QENS spectra of the anionic POPG membrane (a) pure and (b) with 2 mol % RS30 at $Q=1.2$ Å$^{-1}$ using the scattering law given in Eq. 3. (c) Variation of HWHM of the Lorentzian corresponding to lateral motion, $\Gamma_{lat}$, with $Q^2$ for the POPG membrane in absence and presence of 2 mol % RS30. (d) Lateral diffusion coefficient, $D_{lat}$, for the POPG membrane with and without RS30.

Eq. 3 was employed to model the membrane dynamics in both the absence and presence of RS30. The equation successfully describes the QENS spectra for both systems. Representative fitted spectra for POPG membranes without and with 2 mol% RS30 at $Q=1.2$ Å$^{-1}$ are shown in Fig. 9(a) and 9(b), respectively. To gain quantitative insight into the underlying dynamical processes, the extracted fit parameters were analyzed as a function of $Q$. Among these, the lateral motion of lipids is of particular importance, as it underpins various physiologically critical processes, including cell signaling, membrane trafficking, protein localization and activity, and protein–protein interactions. The half-width at half-maximum (HWHM) values of the Lorentzian associated with lateral lipid diffusion is shown in Fig. 9(c). Strikingly, the presence of RS30 significantly restricts lipid diffusion: the HWHM values decrease upon RS30 addition, indicating a reduction in lipid mobility. For both the membrane systems, HWHMs exhibit a linear dependence on $Q^2$, passing through the origin. This behavior confirms that the lipid lateral dynamics follow Fickian continuous diffusion, with $\Gamma_{lat}=D_{lat}Q^2$. The lateral diffusion coefficient $D_{lat}$ was determined using least-squares fitting, with the fit lines shown in Fig. 9 (c). As illustrated in Fig. 9 (d), for pure POPG membranes, $D_{lat}$ is found to be $(10.7\pm0.2)\times10^{-7}$ cm$^2$/s, whereas incorporation of 2 mol% RS30 reduces it by approximately 46 %, yielding $D_{lat}=(5.8\pm0.5)\times10^{-7}$ cm$^2$/s. These findings strongly suggest that RS30 acts as a membrane stiffening agent, significantly hindering lateral lipid diffusion, likely through specific interactions with the polar headgroups of the anionic lipids.

We summarized our findings with a schematic (Fig. 10) showing a representation of the antimicrobial action mechanism of RS30. The figure highlights the selective interaction of RS30 with model bacterial membranes composed of anionic phospholipid (PG), while showing negligible binding to model mammalian membranes composed of zwitterionic lipid (PC). Upon binding to the bacterial membrane, RS30 undergoes a conformational transition from a random coil to an α-helical structure, a change not observed in the mammalian membrane. Interactions of RS30 with the bacterial lipid bilayer disrupt the lipid packing. These perturbations result in membrane depolarization without substantial permeabilization. Furthermore, RS30 significantly restricts the lateral diffusion of lipid molecules in the bacterial membrane, as revealed by QENS, thereby modulating membrane dynamics and enhancing susceptibility to external stresses. This figure summarizes the antimicrobial action mechanism of RS30.

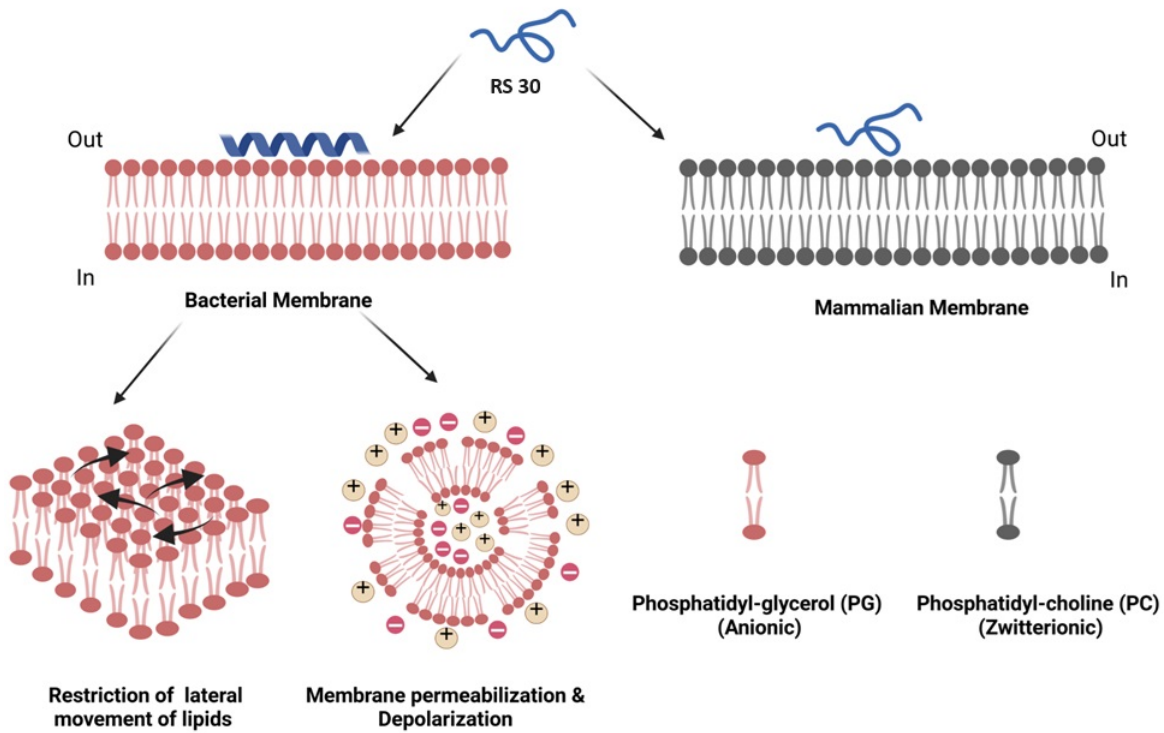

Figure 10: Schematic representation of the selective interaction of RS30 with a model bacterial membrane or mammalian membrane. RS30-membrane interaction causes a conformational transition in the peptide leading to several consequences i.e. restriction of lateral movement of lipids, bacterial membrane permeabilization as well as depolarization.

## CONCLUSIONS

This study offers comprehensive molecular-level insights into the antimicrobial action mechanism of ribosomal protein S30 (RS30) through a combination of microbiological assays and advanced biophysical techniques. Our findings demonstrate that RS30 exhibits potent, time-dependent antibacterial activity against *S. aureus*, primarily mediated by membrane depolarization without substantial membrane permeabilization. This weak perturbation of the membrane by RS30 sets it apart from many classical antimicrobial peptides (AMPs) and suggests alternative action mechanisms such as transient pore formation and/or membrane thinning. These mechanisms provide a plausible explanation for the observed strong depolarization with limited membrane disruption, enabling bacterial death while preserving membrane integrity. Transient water pores, formed due to ionic imbalances, and reduced lipid packing from membrane thinning appear to facilitate this selective depolarization process.

Biophysical studies using model membranes further supported RS30's selective targeting of bacterial membranes. RS30 exhibited strong binding affinity to anionic phospholipid membranes-mimicking bacterial cell envelopes-through electrostatic interactions, while showing minimal interaction with zwitterionic membranes representative of mammalian cells. Upon binding, RS30 underwent a secondary structural transition from a disordered coil to α-helical conformation and induced vesicle aggregation, as confirmed by circular dichroism and dynamics light scattering, respectively. Furthermore, quasielastic neutron scattering measurements reveal that RS30 significantly perturbs the nanoscopic dynamic behavior of lipid molecules in bacterial membranes. Specifically, RS30 reduced the lateral diffusion of anionic lipids, thereby increasing susceptibility to external perturbations-without overtly compromising membrane integrity.

RS30 demonstrated high selectivity by exerting no detectable cytotoxic effects on mammalian cells, including human red blood cells. This selective and semi-lytic mode of action not only enhances its safety profile but also mitigates the risk of inflammatory responses often associated with membrane-lytic peptides. In addition to its therapeutic potential, RS30-derived peptides, due to their high specificity toward bacterial membranes, could also serve as effective infection imaging probes. Their ability to selectively bind to bacterial surfaces without affecting host tissues explains why RS30 based probes have shown promising results in preclinical and clinical studies for *in-situ* diagnosis of infection. It is to be noted that the fragments were reported to not have devastating effect on model membranes or bacterial growth/survival but showed uptake in bacteria [24].

In summary, our study uncovers a unique, semi-lytic mechanism of antimicrobial action by RS30, driven by selective membrane targeting, conformational adaptation, and modulation of lipid dynamics. These findings position RS30 as a promising candidate for next-generation antimicrobial development, particularly in the context of rising antibiotic resistance.

**SUPPORTING INFORMATION:**
The details of data analysis of CD, DLS, ITC and QENS data are provided in the supporting information

# REFERENCES

1. Baker, Rohan T.; Williamson, Nicholas A.; Wettenhall, Richard E. H. The yeast homolog of mammalian ribosomal protein S30 is expressed from a duplicated gene without a ubiquitin-like protein fusion sequence: evolutionary implications. *Journal of Biological Chemistry*, 1996, 271(23): 13549–13555.

2. Kas, Koen; Michiels, Luc; Merregaert, Jozef. Genomic structure and expression of the human fau gene: encoding the ribosomal protein S30 fused to a ubiquitin-like protein. *Biochemical and Biophysical Research Communications*, 1992, 187(2): 927–933.

3. Olvera, Joe; Wool, Ira G. The carboxyl extension of a ubiquitin-like protein is rat ribosomal protein S30. *Journal of Biological Chemistry*, 1993, 268(24): 17967–17974.

4. Perina, Dragutin; Korolija, M.; Hadžija, M. P.; Grbeša, I.; Belužić, R.; Imešek, M.; Morrow, C.; Marjanović, M. P.; Bakran-Petricioli, T.; Mikoč, A.; Ćetković, H. Functional and structural characterization of FAU gene/protein from marine sponge Suberites domuncula. *Marine Drugs*, 2015, 13(7): 4179–4196.

5. Hiemstra, Pieter S., et al. Ubiquicidin, a novel murine microbicidal protein present in the cytosolic fraction of macrophages. *Journal of Leukocyte Biology*, 1999, 66(3): 423–428.

6. Tollin, Maria, et al. Antimicrobial peptides in the first line defence of human colon mucosa. *Peptides*, 2003, 24(4): 523–530.

7. Howell, Scott J., et al. Antimicrobial polypeptides of the human colonic epithelium. *Peptides,* 2003, 24(11): 1763–1770.

8. Brouwer, Carlo P. J. M., et al. Synthetic peptides derived from human antimicrobial peptide ubiquicidin accumulate at sites of infections and eradicate (multi-drug resistant) Staphylococcus aureus in mice. *Peptides,* 2006, 27(11): 2585–2591.

9. Zhou, Xiang, et al. Ribosomal proteins: functions beyond the ribosome. *Journal of Molecular Cell Biology*, 2015, 7(2): 92–104.

10. Lee, Andie S., et al. Methicillin-resistant Staphylococcus aureus. *Nature Reviews Disease Primers*, 2018, 4(1): 1–23.

11. Global antimicrobial resistance and use surveillance system (GLASS) report 2022. Geneva: World Health Organization; 2022. Licence: CC BY-NC-SA 3.0 IGO.

12. Ordonez, Alvaro A., et al. Molecular imaging of bacterial infections: overcoming the barriers to clinical translation. *Science Translational Medicine*, 2019, 11(508): eaax8251.

13. Bassous, N. J., & Webster, T. J. The binary effect on methicillin-resistant *Staphylococcus aureus* of polymeric nanovesicles appended by proline-rich amino acid sequences and inorganic nanoparticles. *Small*, 2019. 15(18), 1804247.

14. Zhang, B., Zhang, M., Lin, M., Dong, X., Ma, X., Xu, Y., & Sun, J. Antibacterial copolypeptoids with potent activity against drug resistant bacteria and biofilms, excellent stability, and recycling property. *Small*, 2022, 18(11), 2106936.

15. Munita, Jose M.; Arias, Cesar A. Mechanisms of antibiotic resistance. *Virulence Mechanisms of Bacterial Pathogens*, 2016, pp. 481–511.

16. Hooper, David C. Fluoroquinolone resistance among Gram-positive cocci. *The Lancet Infectious Diseases*, 2002, 2(9): 530–538.

17. Cao, B., Xiao, F., Xing, D., & Hu, X. Polyprodrug antimicrobials: remarkable membrane damage and concurrent drug release to combat antibiotic resistance of methicillin-resistant *Staphylococcus aureus*. *Small*, 2018, 14 (41), 1802008.

18. Roll, Wolfgang, et al. Infection imaging: focus on new tracers? *Journal of Nuclear Medicine,* 2023, 64(Suppl 2): 59S–67S.

19. Ostovar, A.; Assadi, M.; Vahdat, K.; Nabipour, I.; Javadi, H.; Eftekhari, M.; Assadi, M. *Clinical Nuclear Medicine*, 2013, 38: 413–416.

20. Sathekge, M.; Garcia-Perez, O.; Paez, D.; El-Haj, N.; Kain-Godoy, T.; Lawal, I.; Estrada-Lobato, E. *Annals of Nuclear Medicine*, 2018, 32: 54–59.

21. Welling, M. M.; Nibbering, P. H.; Paulusma-Annema, A.; Hiemstra, P. S.; Pauwels, E. K.; Calame, W. *Journal of Nuclear Medicine*, 1999, 40: 2073–2080.

22. Teixeira, Vitor; Feio, Maria J.; Bastos, Margarida. Role of lipids in the interaction of antimicrobial peptides with membranes. *Progress in Lipid Research*, 2012, 51(2): 149–177.

23. Sharma, Veerendra K., et al. Effect of antimicrobial peptide on the dynamics of phosphocholine membrane: role of cholesterol and physical state of bilayer. *Soft Matter*, 2015, 11(34): 6755–6767.

24. Bhatt Mitra, Jyotsna, et al. Ubiquicidin-derived peptides selectively interact with the anionic phospholipid membrane. *Langmuir*, 2019, 36(1): 397–408.

25. Mizoue, Laura S.; Tellinghuisen, Joel. The role of backlash in the "first injection anomaly" in isothermal titration calorimetry. *Analytical Biochemistry*, 2004, 326(1): 125–127.

26. ITC Data Analysis in Origin, Tutorial Guide Version 5.0; *MicroCal, LLC*: Northampton, MA, 1998.

27. Arnold, Owen, et al. Mantid—Data analysis and visualization package for neutron scattering and μSR experiments. Nuclear Instruments and Methods in Physics Research Section A: Accelerators, Spectrometers, *Detectors and Associated Equipment*, 2014, 764: 156–166.

28. Waghu, Faiza Hanif, et al.CAMPR3: a database on sequences, structures and signatures of antimicrobial peptides. *Nucleic Acids Research*, 2016, 44(D1): D1094–D1097.

29. Nguyen, Leonard T.; Haney, Evan F.; Vogel, Hans J. The expanding scope of antimicrobial peptide structures and their modes of action. *Trends in Biotechnology*, 2011, 29(9): 464–472.

30. Wimley, William C. Describing the mechanism of antimicrobial peptide action with the interfacial activity model. *ACS Chemical Biology*, 2010, 5(10): 905–917.

31. Huang, Yibing; Huang, Jinfeng; Chen, Yuxin. Alpha-helical cationic antimicrobial peptides: relationships of structure and function. *Protein & Cell*, 2010, 1: 143–152.

32. Yount, Nannette Y., et al. Unifying structural signature of eukaryotic α-helical host defense peptides. *Proceedings of the National Academy of Sciences*, 2019, 116(14): 6944–6953.

33. Koynarev, V. R., Nader, M. L., Borgos, K. K. A., Cezar, H. M., Porcar, L., Narayanan, T.& Lund, R. Beyond Structural Pores: Transient Permeabilization of Lipid Membranes by Antimicrobial Peptides. *ChemRxiv*. 2025; doi:10.26434

34. Jean-François, F., Elezgaray, J., Berson, P., Vacher, P., & Dufourc, E. J. Pore formation induced by an antimicrobial peptide: electrostatic effects. *Biophysical journal,* 2008, 95(12), 5748-5756.

35. Ye, G., Gupta, A., DeLuca, R., Parang, K. and Bothun, G.D., Bilayer disruption and liposome restructuring by a homologous series of small Arg-rich synthetic peptides. *Colloids and Surfaces B: Biointerfaces*, 2010. 76(1), pp.76-81.

36. Sato, H. and Feix, J.B Peptide–membrane interactions and mechanisms of membrane destruction by amphipathic α-helical antimicrobial peptides. *Biochimica et Biophysica Acta (BBA)-Biomembranes*, 2006. 1758(9), pp.1245-1256.

37. Lombardi, Lucia; Stellato, M. I.; Oliva, R.; Falanga, A.; Galdiero, M.; Petraccone, L.; D'Errico, G.; De Santis, A.; Galdiero, S.; Del Vecchio, P. Antimicrobial peptides at work: Interaction of myxinidin and its mutant WMR with lipid bilayers mimicking the P. aeruginosa and E. coli membranes. *Scientific Reports*, 2017, 7(1): 44425.

38. Domingues, Tatiana M., et al. Interaction of the antimicrobial peptide gomesin with model membranes: A calorimetric study. *Langmuir*, 2013, 29(27): 8609–8618.

39. Wadhwani, Parvesh, et al. Antimicrobial and cell-*penetrating* peptides induce lipid vesicle fusion by folding and aggregation. *European Biophysics Journal*, 2012, 41: 177–187.

40. Lombardi, Lucia, et al. Antimicrobial peptides at work: Interaction of myxinidin and its mutant WMR with lipid bilayers mimicking the P. aeruginosa and E. coli membranes. *Scientific Reports*, 2017, 7(1): 44425.

41. Karmakar, Sanat; Maity, Pabitra; Halder, Animesh. Charge-driven interaction of antimicrobial peptide NK-2 with phospholipid membranes. *ACS Omega*, 2017, 2(12): 8859–8867.

42. Oliva, Rosario, et al. On the microscopic and mesoscopic perturbations of lipid bilayers upon interaction with the MPER domain of the HIV glycoprotein gp41. *Biochimica et Biophysica Acta (BBA) - Biomembranes*, 2016, 1858(8): 1904–1913.

43. Leontiadou, Hari; Mark, Alan E.; Marrink, Siewert J. Antimicrobial peptides in action. *Journal of the American Chemical Society*, 2006, 128(37): 12156–12161.

44. Park, Sangho, et al. Structural study of novel antimicrobial peptides, nigrocins, isolated from Rana nigromaculata. *FEBS Letters*, 2001, 507(1): 95–100.

45. Cantisani, Marco, et al. Structural insights into and activity analysis of the antimicrobial peptide myxinidin. *Antimicrobial Agents and Chemotherapy*, 2014, 58(9): 5280–5290.

46. Gesell, Jennifer; Zasloff, Michael; Opella, Stanley J. Two-dimensional 1H NMR experiments show that the 23-residue magainin antibiotic peptide is an α-helix in dodecylphosphocholine micelles, sodium dodecylsulfate micelles, and trifluoroethanol/water solution. *Journal of Biomolecular NMR*, 1997, 9: 127–135.

47. Sharma, Veerendra Kumar, et al. Curcumin accelerates the lateral motion of DPPC membranes. *Langmuir,* 2022, 38(31): 9649–9659.

48. Gupta, Jyoti, et al. Microscopic diffusion in cationic vesicles across different phases. *Physical Review Materials*, 2022, 6(7): 075602.

49. Singh, P., Sharma, V. K., Singha, S., García Sakai, V., Mukhopadhyay, R., Das, R., & Pal, S. K.Unraveling the role of monoolein in fluidity and dynamical response of a mixed cationic lipid bilayer. *Langmuir*, 2019, 35(13), 4682-4692.

50. Sharma, V. K., Srinivasan, H., Gupta, J., & Mitra, S. Lipid lateral diffusion: mechanisms and modulators. *Soft Matter*, 2024 20(39), 7763-7796.

51. Sharma, V. K., & Mamontov, E. Multiscale lipid membrane dynamics as revealed by neutron spectroscopy. *Progress in Lipid Research*, 2022, 87, 101179.

52. Mitra, S., Sharma, V. K., Garcia-Sakai, V., Orecchini, A., Seydel, T., Johnson, M., & Mukhopadhyay, R.Enhancement of lateral diffusion in catanionic vesicles during multilamellar-to-unilamellar transition. *The Journal of Physical Chemistry B*, 2016, 120(15), 3777-3784.

53. Sharma, V. K., Gupta, J., Mitra, J. B., Srinivasan, H., Sakai, V. G., Ghosh, S. K., & Mitra, S. The physics of antimicrobial activity of ionic liquids. *Journal of Physical Chemistry Letters*, 15(27), 2024, 7075-708.